\documentstyle[12pt,a4wide]{article}
\begin{document}
\newcommand{\be} {\begin{equation}}
\newcommand{\ee} {\end{equation}}
\newcommand{\ba} {\begin{eqnarray}}
\newcommand{\ea} {\end{eqnarray}}
\newcommand{\e} \epsilon
\newcommand{\la} \lambda
\newcommand{\La} \Lambda 

\author{B. Boisseau\thanks{E-mail :
boisseau@celfi.phys.univ-tours.fr}\\
\small Laboratoire de Math\'ematiques et Physique Th\'eorique\\
\small CNRS/UPRES-A 6083, Universit\'e Fran\c{c}ois Rabelais\\
\small Facult\'e des Sciences et Techniques\\
\small Parc de Grandmont 37200 TOURS, France}
\title{\bf Dynamics of a relativistic Rankine vortex for a two-constituent 
superfluid in a weak perturbation of cylindrical symmetry}
\date{}
\maketitle

\begin{abstract}

From a recent study of a stationary cylindrical solution for a relativistic
two-constituent superfluid at low temperature limit, we 
propose to specify this solution under the form of a
 relativistic generalisation
of a Rankine vortex (Potential vortex whose the core has a solid body
 rotation). Then we establish the dynamics of the central line of this
 vortex by supposing that the deviation from the cylindrical configuration
 is weak in the neighbourhood of the core of the vortex. In ``stiff''
material the Nambu-Goto equations are obtained.

\end{abstract}

\section{Introduction}

First works in phenomenological relativistic superfluidity were
undertaken by Rothen \cite{12}, Dixon \cite{13} and Israel \cite{14, 15}.
The framework of this study is the two-constituent relativistic superfluid
dynamics derived by the convective variational approach. This description,
a specialisation of a general formalism developed by Carter \cite{1, 2} ,
can be considered as a relativistic extention of the standard non dissipative
Landau superfluid dynamics \cite{3} and give the same result \cite{4} as
that of Khalatnikov and Lebedev \cite{6}.

The purpose of this work is to find the relativistic dynamics of the central
line of a vortex in a ``cool'' superfluid when it deviates weakly of the 
stationarity and of the cylindrical symmetry. Carter and Langlois \cite{7,8}
have introduced the denomination cold superfluid at zero temperature and
cool superfluid at low temperature when the only excitations are the phonons
(no rotons).

In a cool superfluid, the precedent authors \cite{8} find the generic vortex
solution (static cylindrically symmetric solution) in Minkowski space. In  this
solution the irrotational covector $\mu_\nu$ which describes the superfluid
constituent becomes spacelike near the center of the vortex. If we attribute
some physical reality to this covector, we must cure this pathological
 situation. One means is to prevent the superfluid constituent to have
access to the neighbourhood of the center of the vortex. This can be obtained
by imposing a constraint on the integrals of motion of the vortex solution.The
second constituent represented by the entropy current vector $s^\sigma$ which
contains all the excitations of the superfluid,posseses the motion of a solid 
body. So the current $s^\sigma$ 
becomes spacelike far of the center of the vortex. This problem has been solved
\cite{8} by a natural cut off radius of the vortex. This situation incites us
 to specify the static cylindrical solution under the form of relativistic
 generalisation of a Rankine vortex \cite{9} (potential vortex whose the core
has a solid body rotation)

As we shall see in the following we can suppose that the core of this vortex
is thin and contains all the excitations. This situation is not without
reminding us superconductor of type II where the normal conductor is confined
in the tubes of magnetic flux. However let us note that it is not a real
 separation of the excitations from the superfluid part since this feature
 arrives in the particular structure of a vortex.

Once defined  the generalisation of the Rankine vortex,
we shall suppose that the central line of this vortex has a small deviation
 from the straight configuration. So we adopt the assumption that
 locally on the boundary of
 the thin core of the vortex, $\mu_\nu$ at the exterior and $s^\sigma$ in the
 interior are those describing the straight vortex.Then, expanding the
 equations of motion in function small parameters caracteristic of the
 curvature and the thickness of the core, the lower order which is the equation of the straight vortex, is automatically verified.The following order gives
 approximatly the dynamical equation of the central line. This method has
been transposed from a study of the dynamics of a  self-gravitating thin
cosmic string \cite{BCL}.

The plan of the paper is the following. In section 2 and 3, we shall review
the essential ingredients of the cool superfluid \cite{7} and of its
stationnary and cylindrical solution \cite{8} which are the base of this work. 
In section 4 we define the generalisation of a Rankine vortex. In
section 5 we obtain the dynamical equation of the central line. 
In stiff matter it is reduced to the Nambu-Goto dynamics. Finally in
section 6 we give a conclusion.

\section{Cool superfluid}

The equation of the relativistic  superfluid mecanics in the non dissipative
limit can be derived \cite{4,7} from a lagrangian 
${\cal L}(s^\sigma,\mu_\sigma)$, where $s^\sigma$ is the entropy current 
and $\mu_\nu$ is the superfluid  momentum covector which is the gradient

\begin{equation}
\label{2.1}
\mu_\sigma=\hbar\nabla_\sigma\varphi
\end{equation}

of the superfluid phase scalar $\varphi$. The variation of ${\cal L}$,

\begin{equation}
\label{2.2}
d{\cal L}=\Theta_{\sigma}ds^{\sigma}-n^{\sigma}d\mu_\sigma ,
\end{equation}
yields the thermal momentum covector $\Theta_\sigma$ associated with
the the entropy current $s^\sigma$ and the particle current $n^\sigma$
associated with the superfluid momentum covector $\mu_\sigma$.

The equations of motion are (\ref{2.1}) and

\begin{equation}
\label{2.3}\\
s^{\rho}(\nabla_\rho\Theta_{\sigma}-\nabla_{\sigma}\Theta_{\rho})=0,\\
\end{equation}
\begin{equation}
\label{2.4}\\
\nabla_{\rho}n^\rho=0,\\
\end{equation}
\begin{equation}
\label{2.5}\\
\nabla_{\rho}s^\rho=0.\\
\end{equation}

The energy-momentum tensor corresponding to the lagrangian ${\cal L}$
has the form
\begin{equation}
\label{2.6}
T^{\rho\sigma}=n^\rho\mu^\sigma+s^\rho\Theta^{\sigma}+\Psi g^{\rho\sigma},
\end{equation}
where
\begin{equation}
\label{2.7}
\Psi={\cal L}-\Theta_\rho s^\rho
\end{equation}
is the pressure function. Equations (\ref{2.1}),(\ref{2.3}),(\ref{2.4}) and
(\ref{2.5}) entail the conservation of the energy-momentum tensor 
$T^{\rho\sigma}$.

The lagrangian ${\cal L}$ is built with the three available scalar quantities
\begin{equation}  
\label{2.8}
c^2s^2=-s_\rho s^\rho,
\end{equation}
\begin{equation}
\label{2.9}
c^2y^2=-\mu_\rho s^\rho ,
\end{equation}
\begin{equation}
\label{2.10}
c^2\mu^2=-\mu_\rho\mu^\rho .
\end{equation}
The variables $n^\rho$ and $\Theta_\rho$ defined by (\ref{2.2}) can be
written in terms of $\mu_\rho$ and $s^\rho$ by
\be
\label{2.11}
n^\rho=\Phi^2 (\mu^\rho-As^\rho)\quad ,\quad \Theta_\rho=\Phi^2 (Ks_\rho+A\mu_\rho)
\ee
where $\Phi^2$, $K$ and $A$ are given by
\be
\label{2.12}
c^2\Phi^2=2\frac{\partial {\cal L}}{\partial \mu^2}\quad , \quad c^2\Phi^2K=
-2\frac{\partial {\cal L}}{\partial s^2}\quad , \quad c^2\Phi^2A=-\frac{\partial {\cal L}}{\partial y^2}.
\ee
The masse current is defined \cite{10} as
\be
\label{2.13}
j^\rho=j^{\rho}_{N}+j^{\rho}_{S}
\ee
where
\be
\label{2.14'}
j^{\rho}_{N}=-m\Phi^2As^\rho
\ee
is interpreted, according to the usual terminology, as the normal mass
current and
\be
\label{2.15}
j^{\rho}_{S}=m\Phi^2\mu^\rho
\ee
as the superfluid current.

In the cool regime in which the entropy is associated just to phonons (not
rotons) the lagrangian is given \cite{7} by
\be
\label{2.16}
{\cal L}=P(\mu^2)-3\psi (s^2, \mu^2, y^2).
\ee
$P$ represents the pressure in the cold limit in which the entropy vanishes
and $\psi$ represents the general pressure of the phonon gaz which is
given by
\be
\label{2.17}
\psi=\tilde{\hbar}\frac{1}{3}c^{-\frac{1}{3}}_s(c^2s^2+(c^2_s-c^2)\frac{y^4}{\mu^2})^{\frac{2}{3}}
\ee
with $\tilde{\hbar}\simeq 0.99\hbar$ and where $c_s$ is the cold sound speed 
determined by $P$, according to
\be
\label{2.18}
\frac{c^2}{c^2_s}=\mu\frac{d\mu}{dP}\frac{d^2P}{d\mu^2}.
\ee

\section{Sationary and cylindrically symmetric solutions}

In a Minkowski space equiped with cylindrical coordinates
\be
\label{3.1}
ds^2=-c^2dt^2+dz^2+dr^2+r^2d\phi^2,
\ee
this class of solutions \cite{8} is characterized by helical current
vectors
\be
\label{3.2}
n^\rho=\nu (r)(1 , v(r) , 0 , \omega (r)),
\ee
\be
\label{3.3}
s^\rho= \sigma (r)(1 , V , 0 , \Omega).
\ee
These currents have no radial components.The quantities $v(r)$ and $\omega (r)$
are respectively the translation velocity and the angular velocity of the
particle current. They are radially dependent.  Whereas the homologous
 quantities $V$ and $\Omega$ of the entropy constituent
are constant.This last requirement express that the ``normal'' constituent
has a rigid motion. This condition prevent dissipation by inevitable
small vicosity which, for non-rigid motion, would appear in the ``normal'' 
flow of a realistic model.

The class of such helical flow includes the more usual subclass of circular
flow for which the longitudinal translation velocities $v$ and $V$ are zero.

The resolution of the equations gives in addition to $\Omega$ and $V$, four other
constants of motion:
\be
\label{3.4}
\overline{\Theta}=\overline{k}^\rho\Theta_\rho
\ee
with
\be
\label{3.5}
\overline{k}^\rho=\sigma^{-1}s^\rho.
\ee
$\overline{\Theta}$ is interpreted as an effective temperature measured in the
corotating frame of the ``normal'' flow.

The components of the superfluid momentum covector
\be
\label{3.6}
\mu_\sigma=(-E , L , 0 , M),
\ee
where the three constants $E$, $L$, $M$ are respectively interpreted
as the energy, the longitudinal momentum and the angular momentum per
particle. So there are six integral constants $\overline{\Theta}$, $V$,
$\Omega$,$E$, $L$ and $M$ which is sufficient to solve the problem which
depends in reality of six independent field components. Let us note that
by Lorentz transformation along $z$ we can choose $L=0$ or $V=0$.

The scalar quantities $\mu$, $s$, $y$ are calculated from (\ref{3.3})
and (\ref{3.6}):
\be
\label{3.7}
c^2\mu^2=\frac{E^2}{c^2}-L^2-\frac{M^2}{r^2},
\ee
\be
\label{3.8}
s^2=\sigma^2(r)(1-\frac{V^2}{c^2}-\frac{\Omega^2r^2}{c^2}),
\ee
\be
\label{3.9}
c^2y^2=\sigma (r)(E-VL-\Omega M),
\ee
and in the cool regime the results recorded in section 2 allow to obtain
an explicit solution \cite{8}.

\section{Relativistic vortex of Rankine}

If a physical meaning is attributed to various currents and momenta $n^\mu$
, $s^\mu$, $\mu_\nu$, $\Theta_\nu$, introduced in this theory, they
must never become spacelike. However in the stationary cylindrical symmetric
solution the expressions (\ref{3.7}) and (\ref{3.8}) show that $\mu_\nu$
and $s^\nu$ can become spacelike for different domains of the radial
coordinate $r$.

For $s^\mu$ the validity of the solution must be limited to a finite
range of $r$. This question has been treated by Carter and Langlois \cite{8}
in order to can describe arrays of vortices in neutron stars. 

For $\mu_\nu$ the question is also important since the momentum becomes
spacelike in the neighbourood of $r=0$ which is the center of the vortex.

In the non relativistic theory, the rotation of the superfluid can be
satisfactorily explained by assuming that it is threaded by a series of
parallel straight vortex lines since there is no principle of limited
velocity in Newtonian theory. (There is of course a physical limit which
 depends of the effective ``thickness of the line'' but this is not
a question of principle.)

On contrary, in the preceeding relativistic theory, the stationary cylindrical
solution cannot competly describe a vortex without some restriction, since, 
for any given solution, the relation (\ref{3.7}) determines a radius
$r=a_0$ by
\be
\label{4.1}
\frac{E^2}{c^2}-L^2-\frac{M^2}{a^2_0}=0
\ee
inside which $\mu_\rho$ becomes spacelike.Therefore the relativity imposes
a condition which does'nt exist in non relativistic theory of superfluid.

It is interesting to have an estimation of the minimum radius $a_0$ in
HeII. We suppose that we have a not too fast circular motion ($L=0$) and
that the vortex is quantified ($M=n\hbar$). From (\ref{4.1}), for one
quantum we have

\be
\label{4.2}
a_0=\frac{Mc}{E}\simeq\frac{\hbar c}{4mc^2}\simeq 5.10^{-15}\,{\mathrm cm}
\ee

There is experimental evidence \cite{11} that the typical core radius $r_0$
of a vortex in HeII is about $10^{-8}$cm. So the limit $a_0$ is probably
never reached.

In the frame of this purely classical relativistic fluid theory, the most
natural way to overcome the theoretical difficulty of the spacelike
current is to suppose that the normal part of the fluid ($s^\rho$, $\Theta_\rho$) is concentrated in the neighbourhood of the center and constitutes a
solid core whereas the the superfluid part ($n^\rho$, $\mu_\rho$) is rejected
around the core. This constitutes a relativistic generalisation of a
Rankine vortex in which the core has a rigid body rotation and the outside
has a circular potential flow.

Such a solution looks as it separates the normal component inside the core
from the superfluid component outside the core. This seems a contradiction
with the usual interpretation where it is stated that the two fluids
cannot be physically separeted. But in our case the fluids are not physically
separated since they  are binded in the same vortex whose the configuration
is imposed by the relativity. This configuration can be compared to the
situation where two reservoirs are connected by a superleak. 
A superleak is a channel through
which the superfluid can flow, but not the normal fluid.This device can
be interpreted as a separation between the two components since there is only
superfluid in the channel. As we shall see below, the comparison
is still yet more striking. Let us examine the question of the temperature.
B.Carter and D.Langlois attributed an effective temperature
$\overline{\Theta}=\overline{k}^\rho\Theta_\rho$ that is uniform throughout
all the fluid. But in the above configuration, the temperature which is associated to the normal fluid (phonon) would have to concerne only the core whereas
outside the core the temperature would be null. Can such a solution exist
in a stationary system? To answere this question let us look again at the
HeII experiments. We know that a temperature gradient can be set up
between two volumes of bulk HeII provided that they are connected by
a superleak \cite{11}. There is also a gradient of pression
to maintain the equilibrium situation, and we have

\be
\label{4.5'}
\frac{\Delta {\mathrm Pression}}{\Delta {\mathrm Temperature}}={\mathrm entropy\;per\; unit\; of\; volume}
\ee

In a relativistic vortex the relativity acts at the frontier of the core 
as a superleak, it prevents the superfluid to get in the
center of the vortex, similarily the superleak clamps the normal fluid, and
as it will be showed below, at the frontier of the core there is a
 relation like (\ref{4.5'}).

Let us precise the equations of this relativistic Rankine model.The difficulty
is to exclude the superfluid part of the center of the vortex. In general
the lagrangian (\ref{2.16}), (\ref{2.17}) does not allows this, since $\psi$
merges $\mu_\rho$ and $s^\rho$. But there is a relation among 
the constants of motion in which $\mu_\rho$ desapears of $\psi$ :

\be
\label{4.6}
\overline{E}=E-VL-\Omega M=0
\ee

In this case $y^2=0$ and $\psi$ is reduced to

\be
\label{4.7}
\psi=\frac{\tilde{\hbar}}{3}c^\frac{-1}{3}_s (c^2s^2)^\frac{2}{3}
\ee    
If we suppose that $c_s$ is a constant, $\psi$ becomes a fonction of $s$
only.

The lagrangian becomes:
\be
\label{4.8}
{\cal L}=P(\mu^2)-3\psi (s^2)
\ee

where $P(\mu^2)$ is the pression of the cold limit (without excitation).

$P(\mu^2)$ rules the exterior of the vortex which is a cold potential fluid.
Since $c_s$ is supposed constant the relation (\ref{2.18}) gives an expression
for $P$

\be
\label{4.8'}
P=a\mu^{\frac{c^2}{c_s^2}+1}+b,
\ee
where $a$ and $b$ are constants of integration.

The relation (\ref{2.11}) between currents and moments becomes

\be
\label{4.9}
n^\rho=\Phi^2 (\mu^2)\mu^\rho
\ee

where $\Phi^2$ is given from (\ref{2.12}) and (\ref{4.8}) by
\be
\label{4.10}
c^2\Phi^2=2\frac{\partial P(\mu^2)}{\partial \mu^2}
\ee
and $\mu^\rho$ by
\be
\label{4.11}
\mu^\rho=(\frac{E}{c^2}, L, 0, \frac{M}{r^2}).
\ee
The equation of motion
\be
\label{4.12}
\nabla_\sigma (\Phi^2 (\mu^2)\mu^\sigma)=0
\ee
is verified.

The core of the vortex is ruled by the part of the lagrangian
$-3\psi (s^2)$ where $\psi$ is the pression of the excitations (phonon gaz).
Let us note that the relation (\ref{4.6}) imposes a radius $r'_0 \leq a_0$ 
inside which $s^\rho$ is not spacelike. The equality is obtained in the
 circular motion.  The relation between currents and moments is reduced to
\be
\label{4.13}
\Theta_\rho=\Phi^2Ks_\rho
\ee
with
\be
\label{4.14}
c^2\Phi^2K=2\times3\frac{\partial \psi}{\partial s^2}=\frac{4}{3}\tilde{\hbar}{c_s}^{-\frac{1}{3}}c^{\frac{4}{3}}\left( s^2 \right)^{-\frac{1}{3}}
\ee
and
\be
\label{4.15}
s_\rho=\sigma (r) (-c^2 , V , 0 , \Omega r^2)
\ee
The temperature of the core is given from (\ref{3.4}) and (\ref{3.5}) by
\be
\label{4.16}
\overline{\Theta}=\frac{4}{3}\tilde{\hbar}c^{-\frac{1}{3}}_s\sigma^{\frac{1}{3}}c^{\frac{4}{3}} \left( 1-\frac{V^2}{c^2}-\frac{\Omega^2 r^2}{c^2}\right)^{\frac{2}{3}}
\ee
hence $\sigma (r)$ is obtained in this case by
\be
\label{4.17}
\sigma (r)=\frac{c_s}{\left( \frac{4}{3}\tilde{\hbar}\right)^3 c^4}\frac{\overline{\Theta}^3}{ \left( 1-\frac{V^2}{c^2}-\frac{\Omega^2 r^2}{c^2}\right)^2}
\ee
The equations (\ref{2.3}) and (\ref{2.5}) are the equations of motion
in the core which has a rigid motion (velocity $V$ along $z$ and rotation
$\Omega$ around $z$).

In this stationary vortex solution, there is a paradoxical gradient of
temperature at the boundary of the core. The temperature is
 $\overline{\Theta}$ in the core and zero at the exterior.
 This is not normally an equilibrium
situation. But let us examine if it is not possible to have a relation like
(\ref{4.5'}) which characterises an equilibrium in a superfluid in presence
of a superleak.

Let us remarck that
\be
\label{a1}
\frac{\psi}{\overline{\Theta}}=\frac{1}{4}\sigma (r)
\ee
where $\sigma (r)$ is the proper density of entropy in the core. If we
choose on the frontier $r_0$ of the core
\be
\label{a2}
P(\mu^2)=-\psi (s^2)
\ee
we obtain on $r_0$
\be
\label{a3}
\frac{\Delta {\mathrm Pression}}{\Delta {\mathrm Temperature}}=\frac{\psi-P}{\overline{\Theta}-0}=\frac{\sigma}{2}
\ee
where $\frac{\sigma}{2}$ is the mean value of the proper density of entropy
at the frontier of the core. Let us note that (\ref{a2}) can be obtained
from (\ref{4.8'}) by choosing the constant $a$ positive and $b$ negative.
$P(\mu^2)$ is a tension in the neighbourhood of the core.

We think that analogy with a superleak is strong enough to accept the difference of pression
and temperature at the boundary between the two fluids.

\section{Dynamics of long waves along the core of the vortex ($c=1$)}

We consider now a little deformation of the relativistic Rankine vortex
described above. Its central line has an arbitrary shape but so that its
curvature is weak. The section of its core stays circular with a little
radius $r_0$. Its motion is slow. We can imagine waves of small 
amplitude along a straight vortex.

One introduces the local coordinates system $(\tau^A,\rho^a)$ attached to the
central line of the vortex:

\be
\label{5.1}
x^\mu=X^\mu (\tau^A)+\rho^a N^\mu_a (\tau^A)
\ee
The Minkowskian coordinates $x^\mu$ are expressed in function of the two coordinates $\tau^A =(\tau^0 , \tau^3)$ of the world sheet sweeped by the central
line, and of two coordinates $\rho^a=(\rho^1 , \rho^2)$ pointing in a
direction orthogonal to the world sheet along the two orthonormal vectors
$N_a^\mu$. We introduce also polar coordinates:
\be
\label{5.2}
\rho^1=r\cos\phi \quad ,\quad \rho^2=r\sin\phi.
\ee

In this problem we need to define a typical length, choosed unity by
convenience, in order to characterize the spacetime neighbourhood of a
point of the central line in which we shall study the vortex. This typical
length is an estimate of the radius of the vortex. We shall suppose
that the characteristic length $l$ on wich we have a change in the 
tangent plane of the world sheet is large and that the radius of the core
 $r_0$ is small
$$l\gg 1\quad , \quad r_0\ll 1 $$

Let us note that we have two typical lenght $l$ and $r_0$.

The vetors $N_a^\mu$ perpendicular to the world sheet and the tangent vectors
$\frac{\partial X^\mu}{\partial \tau^A}$ have little change on a length unity. We can express
 this fact by
\be
\label{5.5}
N_a^\mu=N_a^\mu (\frac{\tau^A}{l})\quad ,\quad \frac{\partial X^\mu}{\partial \tau^A}=X_{,A}^\mu (\frac{\tau^A}{l})
\ee
hence
\be
\label{5.6}
\partial _A N_a^\mu=\frac{1}{l}N^\mu_{a,A} \quad , \quad \partial _B X^\mu_{,A}=\frac{1}{l}X_{,AB}.
\ee

The metric $g_{\alpha\beta}$ in the system of coordinates$(\tau^A,\rho^a)$
can be expressed by
\be
\label{5.7}
g_{AB}=\gamma_{AB}+2K_{aAB}\rho^a+\left(K_{bA}^D K_{aBD}+\delta_{ab}\omega_A\omega_B\right)\rho^a\rho^b
\ee
\be
\label{5.8}
g_{Ab}=\rho^a\epsilon_{ab}\omega_A
\ee
\be
\label{5.9}
g_{ab}=\delta_{ab}
\ee
wherein we recognise the induced metric of the world sheet
\be
\label{5.10}
\gamma_{AB}=X_{,A}^\mu X_{,B}^\nu\eta_{\mu\nu}=O(\frac{1}{l^0}),
\ee
the extrinsic curvature
\be
\label{5.11}
K_{aAB}=\frac{1}{l}\eta_{\mu\nu}N^\mu_{a,A}X^\nu_{,B}=O(\frac{1}{l})
\ee
and the twist defined by
\be
\label{5.12}
\epsilon_{ab}\omega_A=\frac{1}{l}\eta_{\mu\nu}N^\mu_{a,A}N^\nu_b=O(\frac{1}{l}).
\ee
Some other quantities will be useful:
\be
\label{5.13}
\gamma=\det\gamma_{AB},
\ee
the mean curvature
\be
\label{5.14}
K_a=K_{aAB}\gamma^{AB},
\ee
\be
\label{5.15}
g=\det g_{\alpha\beta}=\gamma D^2
\ee
where
\be
\label{5.16}
D=1+K_a\rho^a+\frac{1}{2}(K_aK_b-K^A_{aB}K^B_{bA})\rho^a\rho^b,
\ee
and
\be
\label{5.17}
g^{AB}=\gamma^{AB}-2K_a^{AB}\rho^a+O(\frac{r_0^2}{l^2})
\ee
In the following we suppose that there is no twist in the deformation so that
the cross terms of the metric cancel.

It is natural to adopt the assumption that in the  neighbourhood 
of the core whose the linear dimentions are some unities of $r_0$, the solution coincide at the lower order with the solution of a stationary
 cylindrical vortex described in the preceding section, that is with
(\ref{3.3}) for the interior and (\ref{3.6}) for the exterior. In
Minkowskian coordinates, they are rewritten 
\be
\label{5.18}
\mu_\nu=(-E , L , -\frac{y}{r^2}M , \frac{x}{r^2}M ),
\ee
\be
\label{5.19}
s^\rho=\sigma (r)(1 , V , -y\Omega , x\Omega) ,
\ee
where $\sigma (r)$ is defined in function of the constant $\overline{\Theta}$
by (\ref{4.17}).

We can choose the coordinates $\tau^A$ on the world sheet so that $\gamma_{AB}$ takes the conformally flat form:
\be
\label{5.20}
\gamma_{AB}=F(\frac{\tau^A}{l})\eta_{AB}
\ee
In a point of the world sheet chosen as origine ($\tau^A=0$) we can fix
$\gamma_{AB}=\eta_{AB}$ so that in the neighbourhood we have
\be
\label{5.21}
F(\frac{\tau^A}{l})=1+\frac{\tau^D}{l} f_D(\frac{\tau^A}{l}).
\ee
The holonomic base $\partial_A$, $\partial_a$ is orthogonal but not
orthonormal so that the identification of the solution with the stationary
cylindrical solution must take into account the norm $\sqrt{F}$ of 
$\partial_A$, hence
\be
\label{5.22}
\mu_\sigma=(-\sqrt{F}E , \sqrt{F}L , -\frac{\rho^2}{r^2}M , \frac{\rho^1}{r^2}M),
\ee
\be
\label{5.23}
s^\sigma=\sigma (r)(\frac{1}{\sqrt{F}} ,\frac{V}{\sqrt{F}} , -\rho^2\Omega , \rho^1\Omega), 
\ee
with
\be
\label{5.24}
\sigma (r)=\frac{\overline{\Theta}^3 c_s}{\left(\frac{4}{3}\tilde{\hbar}\right)^3}\left(1-V^2-r^2\Omega^2\right)^{-2}
\ee
are the solutions expressed in the holonomic base$(\partial_A , \partial_a)$.
When expressed in the orthonormal tetrad $(\frac{\partial_A}{\sqrt{F}} , \partial_a)$ we can see that these solutions coincide with the solutions (\ref{5.18}) (\ref{5.19}) of the cylindrical vortex. 

We shall write the equation of motion (\ref{4.12}) for the exterior and
(\ref{2.3}), (\ref{2.5}) for the interior on the frontier of the core and
expand these equations in power of the small parameter $l^{-1}$;. The zero
order which is the solution of the straight vortex vanish identically. In this 
expansion appears also the second small parameter $r_0$ .

For this purpose it will be useful to rewrite the expression (\ref{5.22})
and (\ref{5.23}) in a more condensate form:
\be
\label{5.25}
\mu_\sigma=(\overline{\mu}_A\sqrt{F}\, , \, \mu_a),
\ee
\be
\label{5.26}
s^\rho=\sigma (r)(\frac{\overline{s}^A}{\sqrt{F}}\, , \, \overline{s}^a),
\ee
where
\be
\label{5.27}
\overline{\mu}_A=(-E , L)\quad ,\quad \overline{s}^A=(1 , V)
\ee
and
\be
\label{5.28}
\mu_a=( -\frac{\rho^2}{r^2}M\, , \, \frac{\rho^1}{r^2}M)\quad ,\quad \overline{s}^a=(-\rho^2\Omega\, , \, \rho^1\Omega).
\ee
$\mu_a$ and $\overline{s}^a$ are respetively of order $r_0^{-1}$ and $r_0$.

We must calculate some quantities.Using (\ref{5.17}), (\ref{5.20}), 
(\ref{5.21}), the contravariant components $\mu^\alpha$ are given by
\be
\label{5.29}
\mu^A=g^{AB}\mu_B=\eta^{AB}\overline{\mu}_B-\frac{1}{2}\eta^{AB}\frac{\tau^D}{l} f_D\overline{\mu}_B-2K_a^{AB}\rho^a\overline{\mu}_B+\ldots
\ee
\be
\label{5.30}
\mu^a=\delta^{ab}\mu_b.
\ee
In (\ref{5.29}) the second term is of order $\frac{\tau^D}{l}\leq\frac{1}{l}$, the third term is of order $\frac{r_0}{l}$. The
ellipse designates smaller terms; we shall follow this convention
below. We can express
\be
\label{5.31}
\mu^2=-g^{\alpha\beta}\mu_\alpha\mu_\beta=-g^{AB}\mu_A\mu_B-\frac{M^2}{r^2}
\ee
in function of the corresponding quantity (\ref{3.7}) of the straight
vortex which will be renamed
\be
\label{5.32}
\mu^2_0=-\eta^{AB}\overline{\mu}_A\overline{\mu}_B-\frac{M^2}{r^2}.
\ee
We obtain
\be
\label{5.33}
\mu^2=\mu^2_0+2K_a^{AB}\rho^a\overline{\mu}_A\overline{\mu}_B+\ldots
\ee
and
\be
\label{5.34}
\Phi^2 (\mu^2)=\Phi^2 (\mu_0^2)+2K_a^{AB}\rho^a\overline{\mu}_A\overline{\mu}_B\frac{\partial \Phi^2}{\partial \mu^2} (\mu_0^2)+\ldots.
\ee
We have also
\be
\label{5.34'}
s^2=-g_{\alpha\beta}s^\alpha s^\beta=\sigma^2 (r)(1-V^2-\Omega^2r^2-\delta^2+\ldots),
\ee
with
\be
\label{5.35}
\delta^2=2F^{-1}K_{aAB}\rho^a\overline{s}^A\overline{s}^B=O(\frac{r_0}{l}).
\ee
We can express (\ref{5.34'}) in function of the corresponding quantity
(\ref{3.8}) of the straight vortex renamed $s_0^2$:
\be
\label{5.36}
s^2=s_0^2+\sigma^2 (r)\delta^2+\ldots.
\ee
It is possible now to expand the momentum (\ref{4.13}), (\ref{4.14}), namely 
$\Theta_\rho=d(s^2)^{-\frac{1}{3}}s_\rho$ with $d=\frac{4}{3}\tilde{\hbar}c_s^{-\frac{1}{3}}$ in the form
\be
\label{5.39}
\Theta_\rho=d\left((s_0^2)^{-\frac{1}{3}}-\frac{1}{3}\sigma^2 (r)\delta^2 (s_0^2)^{-\frac{4}{3}}+\ldots\right)\sigma(r)\left(g_{AB}\frac{\overline{s}^B}{\sqrt{F}}\, ,\, \overline{s}_a\right)
\ee
With
\be
\label{5.40}
g_{AB}\frac{\overline{s}^B}{\sqrt{F}}=\sqrt{F}\eta_{AB}\overline{s}^B+\frac{2}{\sqrt{F}}\rho^aK_{aAB}\overline{s}^B+\ldots              
\ee

We can now expand the equations of motion on the frontier $r=r_0$ of the core
of the vortex. The exterior equation (\ref{4.12}) in coordinates
$(\tau^A , \rho^a)$
\be
\label{5.43}
\frac{1}{FD}\partial_\sigma\left(FD\Phi (\mu^2)\mu^\sigma\right)=0
\ee
gives
\be
\label{5.43'}
\frac{1}{FD}\left(\Phi^2 (\mu_0^2)\eta^{AB}\frac{1}{2l}f_A\overline{\mu}_B+
\Phi^2 (\mu_0^2)K_a\mu_b\delta^{ab}+\frac{\partial \Phi^2}{\partial \mu^2}(\mu_0^2)\overline{K}_a\mu_b\delta^{ab}\right)+\ldots=0
\ee
with
\be
\label{5.44}
\overline{K}_a=2K_a^{CD}\overline{\mu}_C\overline{\mu}_D
\ee
In this derivation we have used that for any function $G(r)$:
$$\partial_a\left(G(r)\delta^{ab}\mu_b\right)=0$$
In the equation (\ref{5.43'}) the first term is of order $\frac{1}{l}$, the second and the third are in $\frac{1}{lr_0}$. So we shall retain only the second 
and third terms in $\frac{1}{lr_0}$ which are the leading terms of the expantion (\ref{5.43'}).

Before to continue this calculation let us have a look on the interior
equations (\ref{2.3}) and (\ref{2.5}) which in coordinates $(\tau^A, \rho^a)$
are written
\be
\label{5.47}
s^\rho(\partial_\rho\Theta_\sigma-\partial_\sigma\Theta_\rho)=0
\ee
\be
\label{5.46}
\frac{1}{FD}\partial_\sigma(FDs^\sigma)=0
\ee
Introducing (\ref{5.39}) and (\ref{5.40})in these equations we can express the expansion of these equations. Calculation which is not useful to reproduce, 
shows that after the terms of zero order in $\frac{1}{l}$ which cancel identically from the result of the straight vortex, the leading terms are $\frac{1}{l}$
and $\frac{r_0}{l}.$ These terms are negligeable before the second and
third terms ( in $\frac{1}{lr_0}$) of equation (\ref{5.43'}). Hence
the interior equations do not contribute at the same order and can be
forgotten in the resolution of this problem. Therefore we are left with
the leading term of the equation (\ref{5.43'}):
\be
\label{5.46'}
\left(\Phi^2 (\mu^2_0)K_a+\frac{\partial \Phi^2}{\partial \mu^2}(\mu^2_0)\overline{K}_a\right)\mu^a=0
\ee
This equation is written on the frontier $r=r_0$ of the core; $\mu^a$ depends
of the polar angle $\phi$. Since the equation (\ref{5.46'}) must be
verified for all the polar angles when we turn around the core, we obtain 
finally the equation:
\be
\label{5.47'}
\Phi^2 (\mu^2_0)K_a+\frac{\partial \Phi^2}{\partial \mu^2}(\mu^2_0)\overline{K}_a=0
\ee
Since the radius $r_0$ of the core is supposed small, this equation
gives a good approximation of the equation of motion of the central line.

The equation (\ref{2.16}) can be written
\be
\label{5.48}
\frac{d\Phi^2}{d\mu^2}=\left(\frac{1}{c^2_s}-1\right)\frac{\Phi^2}{2\mu^2},
\ee
hence the equation (\ref{5.47'}) is written:
\be
\label{5.49}
K_a+\frac{1}{2\mu^2_0}\left(\frac{1}{c^2_s}-1\right)\overline{K}_a=0.
\ee
In the stiff material where $c_s=1$, the central line has the
Nambu-Goto dynamics:
\be
\label{5.50}
K_a=0
\ee
Let us note an interesting form of the equations of motion (\ref{5.47'}):
\be
\label{5.51}
K_a^{AB}S_{AB}=0
\ee
where
\be
\label{5.52}
S_{AB}=\frac{\partial\Phi^2}{\partial\mu^2} (\mu^2_0)\overline\mu_A\overline\mu_B+\frac{1}{2}\Phi^2 (\mu^2_0)\eta_{AB}
\ee

We shall study the linear approximation of the equation (\ref{5.51}).
Let us project the extrinsic curvature $K_{aAB}$ in the Minkowski space:
\be
\label{5.53}
K^\mu_{AB}=-\delta^{ab}K_{aAB}N^\mu_b=\partial_A\partial_BX^\sigma\left(\delta^\mu_\sigma-\eta_{\rho\sigma}\eta^{CD}e^\rho_Ce^\mu_D\right)
\ee
whith
\be
\label{5.54}
e^\rho_C=\frac{X^\rho_{,C}}{\sqrt{F}}.
\ee
For a weak deformation of the cylindrical symmetry ($l\gg1$), a rough  
approximation gives:
\be
\label{5.55}
e^x_C \sim e^y_C \sim 0\quad ,\quad e^t_0 \sim e^z_3 \sim 1
\ee
So, from (\ref{5.53}) the leading terms are tranverse ($i=x,y$):
\be
\label{5.56}
K^i_{AB} \approx \partial_A\partial_BX^i,
\ee
and the equations of motion of the core are approximated by
\be
\label{5.57}
S^{AB}\partial_A\partial_BX^i=0.
\ee
These equations are finally transformed into:
\be
\label{5.58}
A\partial_t^2X+2B\partial_t\partial_zX+C\partial_z^2X=0
\ee
where the constant coefficients are 
\be
\label{5.59}
A=\left(\frac{1}{c_s^2}-1\right)E^2-\mu_0^2
\ee
\be
\label{5.60}
B=\left(\frac{1}{c_s^2}-1\right)EL
\ee
\be
\label{5.61}
C=\left(\frac{1}{c_s^2}-1\right)L^2+\mu_0^2
\ee
with
$$\mu_0^2=E^2-L^2-\frac{M^2}{r_0^2}$$
This equation is of hyperbolic type if $B^2-AC>0$.A necessary but not
sufficient condition for that is
\be
\label{5.63}
c_s>\frac{1}{\sqrt{2}}
\ee
This is a necessary condition to have propagating long waves $l\gg1$ along
the core. By a Lorentz tranformation along $z$ we can transform the
equation (\ref{5.58}) into a classical waves equation. But this can be
easily done by choosing the longitudinal momentum $L=0$. Therefore we can
calculate the velocity of propagating long waves which turn out to be
above the velocity of light:
\be
\label{5.64}
v_{pr}=\left(\frac{E^2-\frac{M^2}{r_0^2}}{(2-\frac{1}{c_s^2})E^2-\frac{M^2}{r_0^2}}\right)^{\frac{1}{2}}\geq 1 .
\ee
It is not possible to have propagating long waves along the core in this model
except in stiff material where the velocity of sound is equal to the
velocity of light. In this case the core behaves like a cosmic string.

If $c_s< \frac{1}{\sqrt{2}}$ the equation becomes elliptic and we can
have a motion of large amplitude, but our choice
of weak pertubation of the cylindrical symmetry does not allow to describe 
such a situation.

\section{Conclusion}

These last years the relativistic vortices in superfluid have been analysed
in the context of a model based on non-linear wave equations of a complex
scalar field. These studies enlight the theoretical similarity between
superfluid vortices and global cosmic strings \cite{BY5, BY6, BY8, BY9, BY10}.

In this work, we started from a relativistic two-component fluid model and by specifying a stationary cylindrical solution \cite{8} in the form of a
relativistic Rankine model, we give the dynamics of the core in the small
deviation of the straight configuration. In this way we establish that in
stiff matter the equations of motion of the core are the Nambu-Goto equations
which are also the equations of the cosmic strings.

Let us note that the purposed model seems to have some agreement with
the physical reality where if the core structure of HeII is unknown, there
is some experimental evidence that the excitations of the normal fluid
tend to cluster near the vortex core \cite{11}.

However it is difficult to understand why the long waves cannot propagate
along the core of the vortex when the velocity of sound is not equal
to $c$. Since in general the equations are elliptic the amplitude of motion
can become large, the method does not allow to study a motion without some
severe boundary conditions.

\end{document}